\documentclass[fleqn,12pt,twoside]{article}
\usepackage{espcrc1}

\usepackage{graphicx}
\usepackage[figuresright]{rotating}

\newcommand{\be}{\begin{equation}}
\newcommand{\ee}{\end{equation}}
\newcommand{\ba}{\begin{array}{ccc}}
\newcommand{\ea}{\end{array}}

\def\log{\textnormal{log}}
\def\det{\textnormal{det}}
\def\exp{\textnormal{exp}}
\def\bea{\begin{eqnarray}}
\def\eea{\end{eqnarray}}

\def\la{\lambda}

\def\om{\omega}

\newcommand{\AmS}{{\protect\the\textfont2
  A\kern-.1667em\lower.5ex\hbox{M}\kern-.125emS}}

\title{Partition function zeros in QCD}

\author{Michaela Oswald\address{The Niels Bohr Institute, Blegdamsvej 17, DK-2100 Copenhagen {\O}}\thanks{Done in collaboration with A.~D. Jackson, C.~B. Lang and K. Splittorff}}
              
\begin{document}

\maketitle

The spectral correlations of the QCD Dirac operator in the 4-dimensional Euclidean box, $1/\Lambda_{QCD} \ll L \ll 1/m_{\pi}$, usually denoted as Leutwyler-Smilga regime \cite{LS}, have been the object of intense study in recent years. They contain valuable information about the chiral properties of the QCD vacuum. Provided that chiral symmetry is spontaneously broken, the above regime can be described either by an effective Lagrangian method \cite{LS} or by means of $\chi$RMT (for a recent review see \cite{VW}). The universal limit in which QCD and $\chi$RMT coincide is the limit 
$
N\to\infty\;, \lambda\to 0\;,  m \to 0,
$
in which the microscopic variables, $\zeta\equiv  2\,N\,\lambda$ and $\mu\equiv  2\,N \,m$, are kept fixed and $N$ is identified as the dimensionless volume.  (Here, $\lambda$ denotes an eigenvalue of the Dirac operator 
and $m$ is the  dimensionless quarkmass parameter.)
A way to get information about phase transitions is to study the scaling behaviour of the zeros of the partition function \cite{YL}. The authors of \cite{LS} noticed a qualitative agreement of the QCD partition function zeros and the averaged Dirac operator eigenvalues for one flavour. We show \cite{JLOS} with a ``trapping argument'' in the context of $\chi$RMT that the zeros and the average eigenvalue positions are indeed intimately connected.
It is also known \cite{VW} that the spectral correlations of QCD follow the predictions of $\chi$RMT only up to a certain energy scale, the Thouless energy $E_c$. Beyond this scale correlations die out. By introducing the concept of normal modes we find an alternative way of studying $\chi$RMT. It enables us to remove certain restrictions imposed by $E_c$ and to use $\chi$RMT to describe QCD beyond the Leutwyler-Smilga regime.

The $\chi$RMT partition function corresponding to $N_f$ flavours, topological sector $\nu$, quarks in the fundamental and gluons in the adjoint representation is given by \cite{VW}
\be
Z_{N}^{N_{f},\nu}(\{m\}) = \int{}DW
\prod\limits_{f=1}^{N_{f}} \det(D + m_{f})\;
\exp\left[-N\,\Sigma^2\,{\rm Tr} (W^{\dag}W)\right]\;.\label{pf}
\ee
Here $2N$ is the size of the Dirac operator matrix $D$, $DW$ is the Haar measure over the Gaussian distributed random matrices $W$ and $\Sigma$ corresponds to the chiral condensate in the chiral limit. Rewriting (\ref{pf})  in terms of the eigenvalues of $W$ it becomes an integral over the joint distribution function $P_{N}^{N_{f}}(\lambda_1,\ldots,\lambda_N; \{m_f\})$. To relate the zeros of the partition function to these eigenvalues, we evaluate the collective maximum of $P_{N}^{N_{f}}$ w.r.t. to the $N$ eigenvalue variables. For the quenched case ($N_f=0$) the collective maximum is given by the zeros of the generalized Laguerre polynomial $L_{N}^{\nu-1/2}\left(N\lambda_i^2\right)$. On the other hand, it is also known that the partition function for one flavour is proportional to the Laguerre polynomial $L_{N}^{\nu}\left(-N m^2\right)$. Using general properties of orthogonal polynomials we find that the $i$th zero of the partition function is always between the $i$th and $(i+1)$st value of the collective maximum of the quenched joint distribution function. As we derived this result for arbitrary $N$ it is also valid on the microscopic scale.

For a better understanding of the relation between the mass zeros of the partition function and the maximum of the quenched distribution function we investigate the fluctuations of the $N$ eigenvalues around their collective maximum. We proceed by analogy with \cite{AJP} where this was done for the non chiral ensembles and make the Gaussian approximation
\bea\nonumber
\log\;P_{N}^{N_f=0,\nu} \approx \log\;P_{N;(0)}^{N_f=0,\nu} +
\frac{1}{2}\delta\lambda_{i}\,C_{ij}\,\delta\lambda_{j}~.
\eea
Here the $\delta\lambda_i$ are the relative eigenvalues w.r.t.~their value at the collective maximum. The eigenfunctions of the matrix $C$ are evaluated at the maximum and are called normal modes. They describe the statistically independent correlated fluctuations of the eigenvalues around their most probable values. The eigenvalue equation is $\sum_{i=1}^{N}C_{ij} \phi_{j}^{(k)} = \om_{k} \phi_{i}^{(k)}$. In the large $N$ limit the eigenfunctions are Chebyshev polynomials $\phi_{i}^{(k)}=\sqrt{2/N}\;U_{2k-1}\left(\la_{i}/2\right)$. The dispersion relation turns out to be linear for all $k$ and $N$,
\be
\om_{k}= -4\,k\,N~.
\ee
This is in contrast to a quadratic dispersion relation for uncorrelated eigenvalues. 

What do these normal modes have to do with the Thouless energy $E_c$? The authors of \cite{JMRSW} found that the presence of $E_c$ means a strong enhancement of the longest wave length modes, whereas the rest are still given by random matrix theory. An enhancement of these modes can easily be incorporated in our Gaussian approximation to mimic the presence of a Thouless energy. As expected the partition function zeros in the microscopic limit, i.e. in the Leutwyler-Smilga regime, remain uneffected. Moreover, it was found in \cite{JMRSW} that removing these soft modes leads to an agreement with RMT on all scales.

The concept of spectral normal modes is generally applicable and not restricted to random matrix theory. In particular it could be studied in lattice QCD. Such a study would provide valuable information about $\chi$RMT correlations in lattice QCD. We expect that almost all of the normal modes are given by $\chi$RMT and the longest wave length modes will depend on the detailed dynamics.

\end{document}